\documentclass[conference]{IEEEtran}
\usepackage[latin9]{inputenc}
\usepackage{units}
\usepackage{algorithm2e}
\usepackage{amsmath}
\usepackage{graphicx}

\makeatletter

\providecommand{\tabularnewline}{\\}

\usepackage{color}
\usepackage{soul}
\usepackage{cite}

\def\BibTeX{{\rm B\kern-.05em{\sc i\kern-.025em b}\kern-.08emT\kern-.1667em\lower.7ex\hbox{E}\kern-.125emX}}

\makeatother

\begin{document}
\title{ABSense: Sensing Electromagnetic Waves on Metasurfaces via Ambient
Compilation of Full Absorption}
\author{C.~Liaskos$^{1*}$, G.~Pirialakos$^{1,2}$, A.~Pitilakis$^{1,2}$,
S.~Abadal$^{3}$, A.~Tsioliaridou$^{1}$, A.~Tasolamprou$^{1}$,\\
 O.~Tsilipakos$^{1}$, N. Kantartzis$^{2}$, S.~Ioannidis$^{1}$,
E.~Alarcon$^{3}$, A.~Cabellos$^{3}$, M. Kafesaki$^{1}$,\\
 A.~Pitsillides$^{4}$, K.~Kossifos$^{4}$, J.~Georgiou$^{3}$,
I.F. Akyildiz$^{4,5}$\\
{\small{}$^{1}$FORTH, Greece, $^{2}$AUTH, Greece, $^{3}$UPC, Barcelona,
Spain, $^{4}$UCY, Nicosia, Cyprus},{\small{} $^{5}$ECE Gatech, USA
}\\
{\small{}$^{*}$Corresponding author email: cliaskos@ics.forth.gr}}
\maketitle
\begin{abstract}
Metasurfaces constitute effective media for manipulating and transforming
impinging EM waves. Related studies have explored a series of impactful
MS capabilities and applications in sectors such as wireless communications,
medical imaging and energy harvesting. A key-gap in the existing body
of work is that the attributes of the EM waves to-be-controlled (e.g.,
direction, polarity, phase) are known in advance. The present work
proposes a practical solution to the EM wave sensing problem using
the intelligent and networked MS counterparts-the HyperSurfaces (HSFs),
without requiring dedicated field sensors. An nano-network embedded
within the HSF iterates over the possible MS configurations, finding
the one that fully absorbs the impinging EM wave, hence maximizing
the energy distribution within the HSF. Using a distributed consensus
approach, the nano-network then matches the found configuration to
the most probable EM wave traits, via a static lookup table that can
be created during the HSF manufacturing. Realistic simulations demonstrate
the potential of the proposed scheme. Moreover, we show that the proposed
workflow is the first-of-its-kind embedded EM compiler, i.e., an autonomic
HSF that can translate high-level EM behavior objectives to the corresponding,
low-level EM actuation commands.
\end{abstract}


\section{Introduction}

Metasurfaces (MS) are highly efficient media for the manipulation
of electromagnetic (EM) waves in custom and even unnatural ways. Wave
attributes such as the direction of reflection, polarization~\cite{yang2016programmable},
and the amount of absorption~\cite{PlasmonicMS}, among others, can
be controlled with unprecedented accuracy. The outstanding principles
MS have been already proven in the microwave, Terahertz (THz), and
optical regimes~\cite{Tasolamprou2019720}, finding their way into
a plethora of applications~\cite{wallace2010analysis,Liaskos:2018:UAS:3289258.3192336,MSSurveyAllFunctionsAndTypes}.

The basis for the MS operation is the Huygens principle, which states
that any EM wavefront can be traced back to a planar distribution
of currents~\cite{MSSurveyAllFunctionsAndTypes}. In that sense,
a MS acts as a canvas of currents comprising: i) passive elements
acting as sub-wavelength antennas receiving impinging waves, i.e.,
inductive current sources, and ii) active elements such as PIN diodes
acting as current flow manipulators. The active elements can be externally-biased,
transforming the currents to follow a planar distribution that corresponds
to a required wavefront. Thus, any overall manipulation of the impinging
EM waves can be attained, e.g., anomalous steering, polarization and
phase alteration, partial attenuation or even full absorption. HSFs
are the intelligent, networked variation of MSes, which include an
embedded nano-network and a gateway~\cite{AccessUPC}. The nano-network
collectively controls the biasing of the active elements in a distributed,
autonomic manner, while the gateway connects the nano-network to the
HSF-external world via a standard protocol (e.g., WiFi, Bluetooth,
etc.).

It is a widespread practice to design MSes (and HSFs) assuming that
the impinging wave attributes are known~\cite{MSSurveyAllFunctionsAndTypes}.
Nevertheless, this is rarely the case in a broad set of applications
where the availability of EM sensing systems becomes a necessity.
Currently, the wave attributes can be sensed via HSF-external systems
and devices (e.g., d-dot sensors)~\cite{Fieldsen0:online}. However,
this approach is not space-granular, since the EM field quantities
are sensed i) on average and ii) at one point near or on the HSF gateway.
Incorporating multiple such sensors in the HSF adds up to its assembly
complexity and overall cost. Moreover, d-EM field sensors are currently
size-able, posing novel minification challenges.

The present study proposes a wave sensing approach that operates without
specialized sensory hardware and exploiting the HSF networking capabilities
instead. The key idea is to sense the attributes of an impinging wave
by fully absorbing it. Full absorption of impinging EM waves is well-studied
MS capability~\cite{MSSurveyAllFunctionsAndTypes}. When the surface
impedance of the HSF is matched perfectly to the impinging wave, its
power dissipates on the passive and active elements, at the same time
serving the dual purpose of identifying the condition for perfect
absorption. Thus, we use the embedded controller network within the
HSF to intelligently iterate over the active element states and obtain
the one yielding maximal dissipated power across the active elements.
Then, we employ a static lookup table (provided by the HSF manufacturer)
that contains the actuator states to achieve full absorption for specific
impinging wave cases. The best matching entry is picked as the estimate
of the impinging wave attributes. Apart from acting as a wave sensing
scheme, the proposed approach is also a form of an ambient EM Compiler~\cite{LiaskosComp},
since the nano-nodes collectively tune the HSF to attain a macroscopic
functionality, i.e., the full absorption of an unknown EM wave.

The remainder of the paper is organized as follows. Section~\ref{sec:related}
surveys the related studies. Section~\ref{sec:proposed} details
the proposed scheme and evaluation via simulations follows in Section~\ref{sec:evaluation}.
The paper is concluded in Section~\ref{sec:conclusion}.

\section{Related Studies}

\label{sec:related}

Metasurfaces are highly effective systems for controlling different
aspects of EM waves (wavefront \cite{Diaz-Rubio:2017}, collimation
\cite{Tasolamprou201423147}, polarization \cite{Niemi20133102},
dispersion \cite{Tsilipakos:2018}, controllable absorption \cite{Wang2018}),
especially if they are tunable thus offering the ability to perform
multiple functionalities and switch between them at will \cite{Tasolamprou2019720}.
In recent years, interest has mainly focused on incorporating voltage-controlled
actuation elements inside each unit cell, thus providing a means of
dynamically tuning the surface impedance of the metasurface to external
EM waves. This is because voltage-controlled elements can be readily
controlled by an external controller, such as an Field Programmable
Gate Array (FPGA), thus allowing for centrally controlling the metasurface
response. The commonly integrated elements are PIN switch \cite{Cui:2014}
or varactor \cite{Huang:2017} diodes, allowing for a broad range
of functionalities ranging from wavefront manipulation, beam splitting,
and polarization control. However, switch diodes allow for obtaining
two discrete states of the surface impedance; it can be extended to
$2^{N}$ when $N$ unit cells are clustered together to allow for
$N$-bit encoding, but $N$ is typically limited to $N=2$ or $3$~\cite{Cui:2014}
since the resulting supercell must be subwavelength to remain in the
metasurface regime. On the other hand, varactor diodes allow for continuous
control but only over the reactive part of the surface impedance,
thus providing the ability to tune the phase of the impinging wave
but not its amplitude.

Recently, it has been shown that having the ability to continuously
control both the reactive and resistive part of the complex surface
impedance leads to maximum freedom over the available functionalities
\cite{Fu:2019}. This can be achieved by integrating elaborate controller
chips in each unit cell that can provide a complex input impedance
($R+jX$). Further enhancing this concept by forming a nano-network
of the controllers defines the HyperSurface (HSF) paradigm~\cite{DBLP:conf/WoWMoM/Liaskos,AccessUPC}.
In a nutshell, the HyperSurface entails the passive meta-atom enhanced
with an actuation module, a computation and communication module for
exchanging data with other cells, and a gateway for communicating
with HSF-external entities, such as receiving cell actuation commands
and diffusing them for propagation within the inter-cell network.
A prominent example of the applications enabled by HSFs are employing
mitigation strategies for fault tolerance~\cite{DBLP:journals/corr/abs-1810-06329}
or deploying programmable wireless environments~\cite{DBLP:conf/WoWMoM/Liaskos,Liaskos2019ADHOC,IEEEcomLiaskos,Liaskos:2018:UAS:3289258.3192336}.

Here, we demonstrate a different possibility enabled by the rich HSF
capabilities. Specifically, we demonstrate sensing of the characteristics,
i.e., direction of incidence (measured by the spherical coordinate
angles $\phi$ and $\theta$) and polarization (TE or TM), of the
impinging wave. Similar functionalities have been demonstrated in
plasmonic metasurfaces incorporating pixelated photodetectors strictly
for polarization detection~\cite{Pelzman20161213} and for orbital
angular momentum detection enabled through the decomposition of the
transmitted waves impinging on a properly designed metasurface~\cite{Chen2018110}.
In our approach we utilize the controllers integrated in the HSF configuration.
In particular, we rely on the fact that by fully absorbing the incident
wave the power dissipated on the controllers is maximized. More specifically,
we monitor the dissipated power on the actuation elements themselves,
thus avoiding extra sensing circuitry which would increase the complexity.
Moreover, the proposed sensory system is autonomic, not requiring
HSF-external computing or communication elements~\cite{Fieldsen0:online}.
Additionally, to be best of the authors' knowledge, the present work
also constitutes the first application of distributed consensus algorithms~\cite{consensus1,consensus2}
as the enabler for the autonomic and collective operation of nano-networks.
In this aspect, we note that related works have studied controlled
flooding~\cite{akram,NANOCOM2015Tsioliaridou}, peer-to-peer~\cite{julien,akram,ICTsioliaridou,NANOCOMlocalizeSDM,al2018lagoon}
and nature-inspired alternatives~\cite{ICCLiaskos}.

It is noted that impressive further capabilities of metasurfaces has
been recently demonstrated, promising novel expansions of the HyperSurface
concept. Estakhri et al have showcased metastructure design processes
able to serve as analog solvers to integral problems~\cite{estakhri2019inverse}.
La Spada et al have proposed design workflows for curvilinear metasurfaces
exerting arbitrary control over surface EM currents~\cite{la2019curvilinear},
as well as near-zero index wires~\cite{la2017near}. Finally, graphene
has also been extensively studied for frequency-tunable THz metasurfaces,
since its conductivity can be dynamically modulated via electrical,
magnetic and optical means~\cite{vakil2011transformation}.

\section{The proposed scheme}

\label{sec:proposed}

\begin{figure}[t!]
\centering{}\includegraphics[width=1\linewidth]{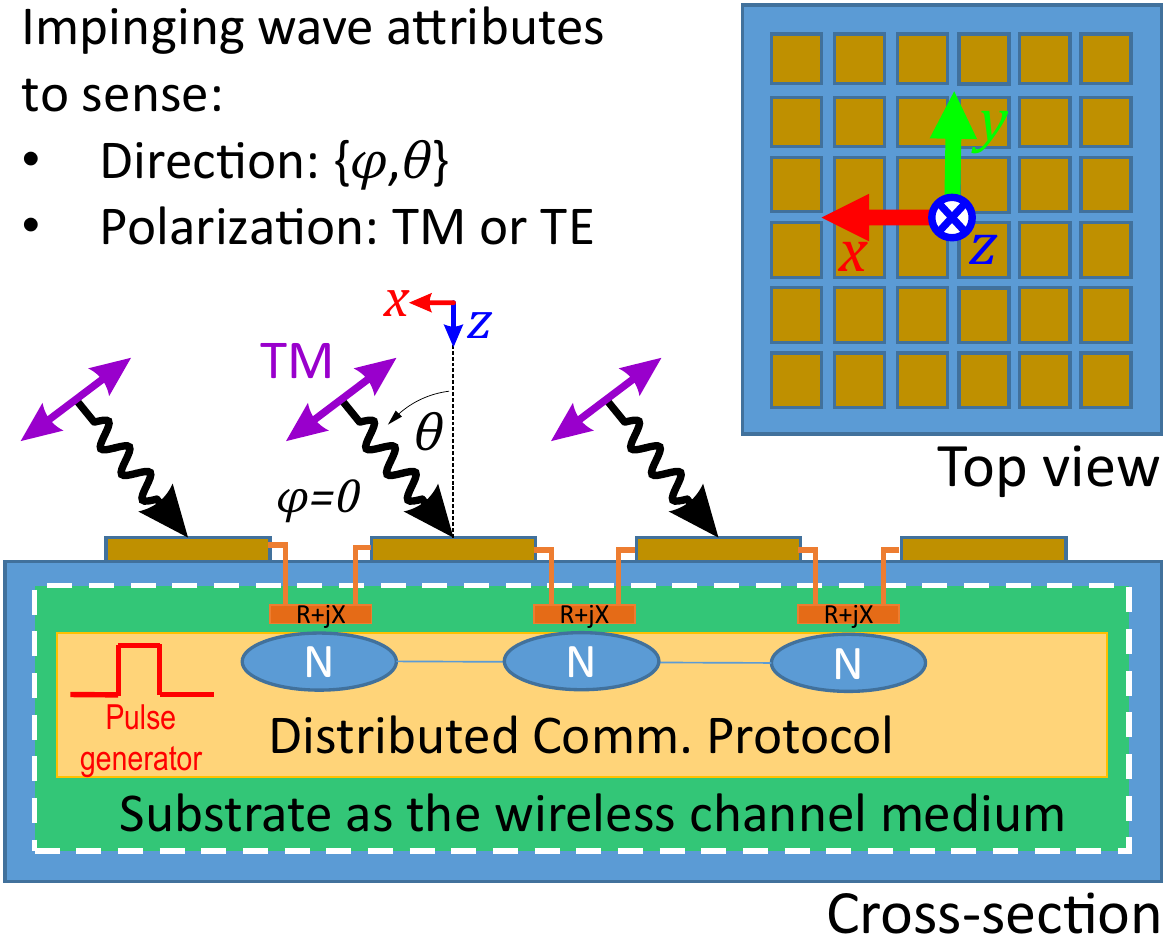} \caption{\label{fig:setup}The generic HyperSurface structure comprising passive
elements (conductive patches and insulating substrate) and active
elements (tunable impedance elements and their networked nano-controllers).
The attributes of the impinging planar wave to be sensed are its direction
of arrival (azimuth and elevation angles expressed in the illustrated
coordinate system) and polarization. }

\end{figure}
We consider a generic HSF architecture as shown in Fig.~\ref{fig:setup}.
From its physical aspect, it comprises the common passive and active
elements of a metasurface. Additionally, it includes a wireless nano-network
embedded within the HSF material, with each nano-node being responsible
for the control of one active element. Notably, the HSF material also
constitutes the wireless channel medium for the communicating nano-nodes~\cite{Tasolamprou2018,Timoneda2018b}.
A distributed communications protocol is employed by the nodes at
the application layer, as described later in this Section. The nano-network
is supplied with power via an energy pulse generator, which also serves
as a rudimentary clock as discussed later. The power pulses can be
of any spectrum, such as microwaves, visible light, etc.~\cite{eharvest}.
A planar wave impinges upon the HSF, with a specific direction and
polarization, as illustrated in Fig.~\ref{fig:setup}. The objective
is to employ the nano-network capabilities in order to detect these
characteristics of the wave, i.e., direction and polarization. The
proposed process executed by the HSF as a whole is denoted as \emph{ABSense}
and formulated in pseudo-code as Algorithm.~\ref{alg:absense}.

\begin{algorithm}[t!]
\caption{\label{alg:absense}The workflow of the proposed ABSense scheme.}
\SetAlgoLined

$Input_{1}$: An impinging planar wave {[}$\phi$, $\theta$, polarization{]}\;

$Input_{2}$: The read-only HSF HashMap $L$: {[}Parameterized EM
Function ID{]}$\to${[}Local Impedance $z${]}\;

$Output$: The sensed wave attributes {[}$\phi^{*}$, $\theta^{*}$,
polarization$^{*}${]}\;

- initialization\;

\For{each mapped value $z$ of HashMap $L$}{

- Set all active elements of the HSF to $z$\;

- Measure the power flow $P_{e}\left(z\right)$ at the active elements
of each node $e$\; }

- Pick $z_{e}^{*}\gets argmax\left\{ P_{e}\left(z\right)\right\} $\;

- Obtain the average $E_{e}\left[z_{e}^{*}\right]$ over all nodes\;

- Pick $f_{id}^{*}\gets argmax\left\{ L:\,z=z_{e}^{*}\right\} $\;

- Return parameters {[}$\phi^{*}$, $\theta^{*}$, polarization$^{*}${]}
of $f_{id}^{*}$\;
\end{algorithm}
The goal of the ABSense scheme is to sense the impinging wave attributes
by: i) detecting the active element configurations that leads to the
full absorption of the wave, and ii) deducing the best matching wave
attributes by reversing a hashmap provided by the HSF manufacturer.
This hashmap, denoted as $L$, corresponds any parameterized EM function
supported by the HSF to the active element state per node $e$ to
achieve it. For instance, an entry of $L$ can be written as:
\begin{equation}
{\scriptstyle f_{1}:\,\text{FULL\_ABSORB\ensuremath{\left(\varphi\text{:}\,30^{\text{o}},\,\theta:\,42^{o},TE\right)}\ensuremath{\ensuremath{\to z_{e}:R_{1}}+j\ensuremath{X_{1}}}}}\label{eq:example}
\end{equation}

Treating $L$ and the impinging wave as inputs, the ABSense workflow
is as follows. Signaled and synchronized by the pulse generator, each
nano-node iterates over its possible active element states. All nodes
move along their iterations in lock-step (detailed below), resulting
each time into a uniform surface impedance across the HSF, which will
eventually match the impinging wave. For each state $z$, a node $e$
obtains a corresponding measurement of power flowing via its active
element, $P_{e}\left(z\right)$. Susceptibility to noise and errors
is taken into account. The $z_{e}^{*}$ state that yields the maximum
$P_{e}$ is communicated to the nano-network as a whole via a distributed
consensus approach, and an average value, $E_{e}\left[z_{e}^{*}\right]$,
is obtained. Finally, using the $L$ map in reverse, each nano-node
obtains the best matching parameterized function and, hence, the best
matching wave attributes as well.

\begin{figure}[h]
\centering{}\includegraphics[width=1\linewidth]{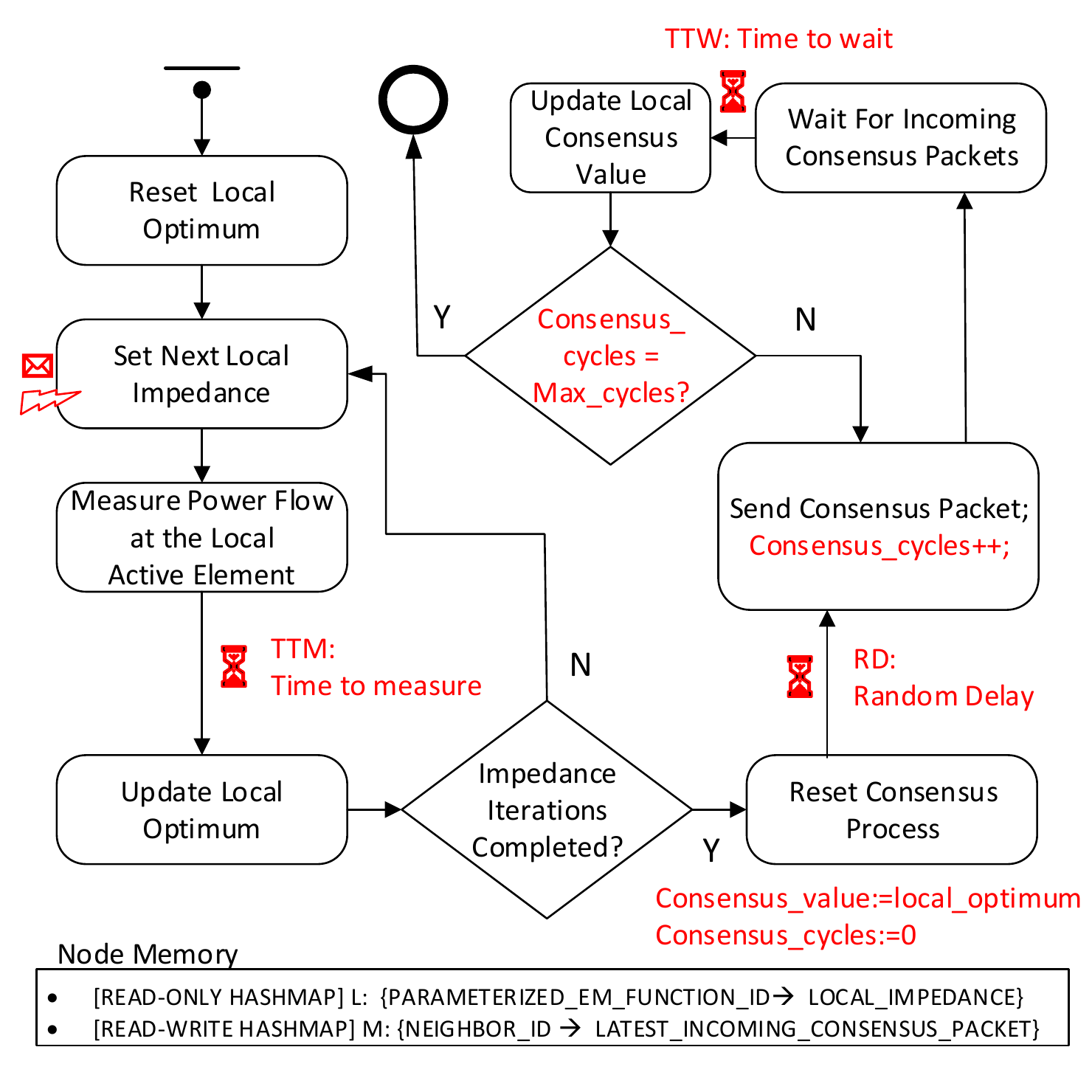} \caption{\label{fig:workflow}The nano-node memory structures and operation
in state-chart form (measurements and consensus sub-processes included). }

\end{figure}
We proceed to detail the operation of each nano-node in Fig.~\ref{fig:workflow}.
It comprises two phases, the iteration over local active element states
(a set of impedance values $\left\{ z_{e}:\,R+jX\right\} $) to detect
the one ($z_{e}^{*}$) yielding the maximum power flow, and the consensus
phase to obtain the $E_{e}\left[z_{e}^{*}\right]$. From another point
of view, this phase is also an ambient compilation of an EM functionality,
since the nano-nodes self-tune the HSF to perform a full EM wave absorption~\cite{LiaskosComp}.
The iteration phase is initialized by resetting the $z_{e}^{*}$ variable.
Subsequently, the nodes are iterating over all their impedance values,
orchestrated by the pulse/power generator (considered to be physically
decoupled from the inter-nanonode packet exchange process). We assume
that the pulse generator emits its energy in the form of pulse sequences
representing integer identifiers of the nodes' impedance values. This
pulse emission and orchestration via impedance value identifier broadcast
process is continuous. The identifiers are broadcast given enough
time for the impedance value to be setup and the power measurement
to be obtained (\textbf{TTM}: time to measure), accounting for clock
drifts and signal processing variations.

The consensus phase then operates as follows. Each node initializes
its consensus value as $E_{e}\left[z_{e}^{*}\right]\gets z_{e}^{*}:\,R+jX$,
and broadcasts it to all nano-nodes in its vicinity. A random delay
(\textbf{RD}) is considered to minimize packet collisions. The broadcast
date has the form of a packet structured as follows:\\
~~~~~~~ %
\begin{tabular}{|c|c|c|}
\hline
SENDER\_ID (8 bit)  & R (32 bits) & X (32 bits)\tabularnewline
\hline
\end{tabular}\\
where the sender ID is an integer identifier of the sending node.
During a fixed time interval (\textbf{TTW}: time to wait), each receiving
node collects incoming packets while also keeping a log of the average
reception power for each one. This information is kept in a hash-map,
$M_{\text{\ensuremath{{\scriptscriptstyle \text{SENDER\_ID}}}}}\to E\left[z_{\text{\ensuremath{{\scriptscriptstyle \text{SENDER\_ID}}}}}^{*}\right]^{*}$,
with sender IDs as keys, as shown in Fig.~\ref{fig:workflow}. Subsequently,
each node obtains a first estimate of $E_{e}\left[z_{e}^{*}\right]$
as:
\begin{equation}
E_{e}\left[z_{e}^{*}\right]\gets E_{e}\left[z_{e}^{*}\right]\cdot w_{e}+\underset{\forall{\scriptscriptstyle \text{SENDER\_ID}}}{\sum}M_{{\scriptscriptstyle \text{SENDER\_ID}}}\cdot w_{{\scriptscriptstyle \text{SENDER\_ID}}}
\end{equation}
where $w_{e}\in\left(0,1\right)$ is the personal weight that the
node gives to its local value, and $w_{{\scriptscriptstyle \text{SENDER\_ID}}}$
is the weight assigned to the incoming consensus packets. We consider
that the $w_{{\scriptscriptstyle \text{SENDER\_ID}}}$ values are
proportional to the reception power of the corresponding incoming
packets, and are normalized to comply to the condition:
\begin{equation}
w_{e}+\underset{\forall{\scriptscriptstyle \text{SENDER\_ID}}}{\sum}w_{{\scriptscriptstyle \text{SENDER\_ID}}}=1
\end{equation}
The consensus process converges iteratively, allowing each node to
obtain the actual $E_{e}\left[z_{e}^{*}\right]$ value~\cite{consensus2}.
The consensus process is allowed to run for a maximum number of send/receive
packet cycles (\textbf{max\_cycles}), upon which it yields the estimated
$E_{e}\left[z_{e}^{*}\right]$. The process then concludes by reversing
$L$ to detect the full EM absorption parameterized function that
best matches the estimated $E_{e}\left[z_{e}^{*}\right]$ and, subsequently,
it returns the corresponding EM function parameters as the most probable
EM wave attributes.

The process as a whole can be immediately restarted, should the HSF
tile be operating in a dedicated sensory role, i.e., to sense EM waves
and inform other HSF units to adapt accordingly. Alternatively, having
obtained the sensory information, the same HSF tile can autonomically
apply a different EM manipulation function, such as steering the sensed
wave towards a direction. The mode of operation can be adapted to
the application scenario at hand.

\section{Evaluation}

\label{sec:evaluation}

\begin{figure}[t!]
\centering{}\includegraphics[width=0.8\linewidth]{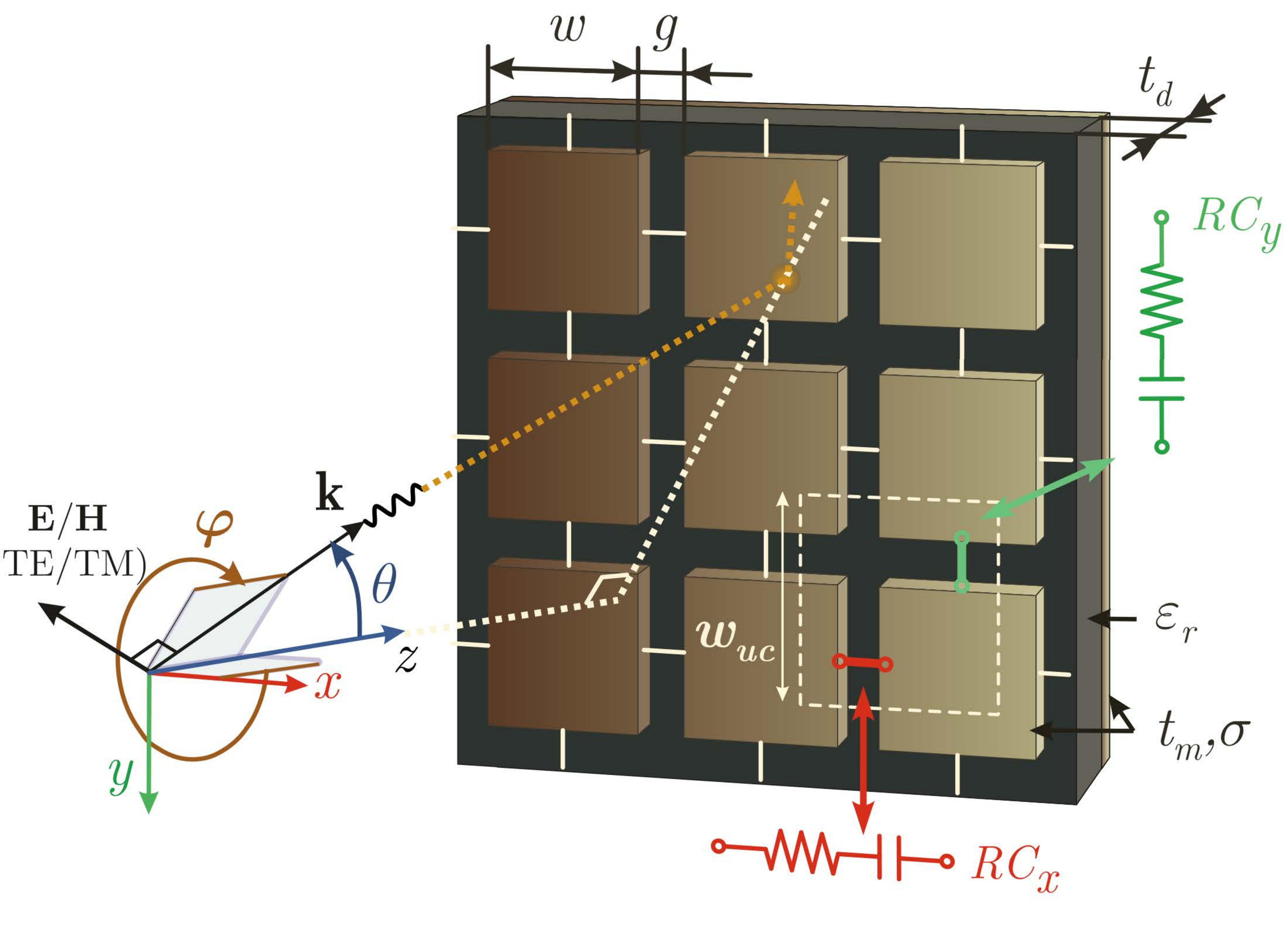}
\caption{\label{fig:Schematic}Metasurface used for ABSense in this work, with
annotation of the main unit-cell parameters and the coordinate systems,
Cartesian ($xyz$) and spherical ($\theta,\varphi)$. Incident wave
polarization is termed TE (or TM) when the $E$-field vector is perpendicular
(or parallel) to the plane of incidence, defined by the $z$-axis
and an azimuth angle $\varphi$.}
\end{figure}

We evaluate the proposed scheme using simulations. The evaluation
combines full wave simulations conducted in CST~\cite{CST} and nano-network
simulations conducted in the AnyLogic platform~\cite{XJTechnologies.2013}.
The full wave simulations produce a dataset of power flow values at
each active element in a specific HSF design (shown in Fig.~\ref{fig:Schematic})
for a variety of impinging wave directions and polarizations, and
for any active element state each time. This dataset is then passed
on to the nano-network simulator which follows the process described
in Section~\ref{sec:proposed}.

The HSF used in this work consists of an array of $30\times30$ square
metallic patches over a thin metal-backed dielectric (Fig.~\ref{fig:Schematic}
illustrates a $3\times3$ sub-part for ease of presentation). The
unit-cell includes two lumped elements modeled as complex-valued $RC$
(resistive and capacitive) loads that connect neighbouring metallic
patches, along the $x$- and $y$-directions. The corresponding unit-cell,
framed with a white border, is designed for 5~GHz operation and its
main parameters are: $w_{uc}=10$~mm, $w_{p}=4.5$~mm, $g=0.5$~mm,
$t_{m}=0.02$~mm, $t_{d}=0.5$~mm, $\varepsilon_{r}=2.2$ and $\sigma=5.8\times10^{7}$~S/m.
For normal incidence, perfect absorption of $x$- and $y$-polarized
plane waves is achieved when $R_{x}=R_{y}\approx1.15$~Ohm and $C_{x}=C_{y}\approx0.99$~pF.

\begin{figure}[t!]
\centering{}\includegraphics[width=1\linewidth]{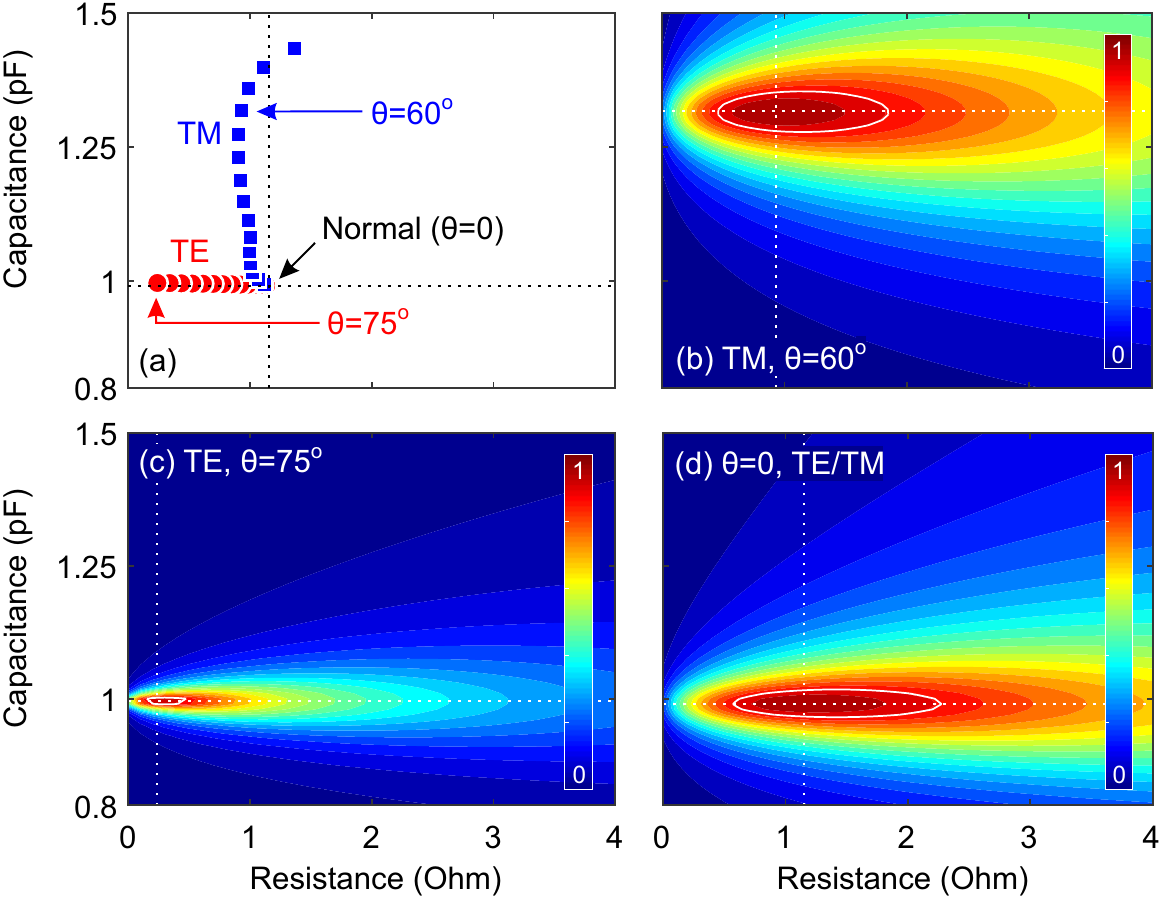}
\caption{\label{fig:Colormaps}(a) $RC$ loads required for perfect absorption
of obliquely incident TE and TM polarized waves in the $xz$ or $yz$
planes, at 5~GHz and for $\theta=0$ to $75^{o}$ with $5^{o}$ steps.
(b)-(d) Absorption coefficients for a few polarization and incidence
direction combinations in the $xz$ plane, as the $RC_{y}$ or $RC_{x}$
are varied for TE or TM polarizations, respectively.}
\end{figure}

Sample visual hash-maps that are used to link the $RC_{xy}$ values
required for perfect absorption with incidence direction and polarization
are depicted in Fig.~\ref{fig:Colormaps}: panel (a) shows the required
$RC_{y/x}$ for TE/TM polarized incidence as the elevation angle $\theta$
varies, for azimuth angle $\varphi=0$, i.e., when $xz$ is the plane
of incidence; panels (b)-(d) show the absorption coefficient for waves
of three different incidence directions and polarizations as the $RC$
values are varied; the white circles mark the regions where the attenuation
coefficient is higher than 0.9 (corresponding to reflection coefficient
of $-10$~dB or lower), and the white crosses mark the $RC$ values
leading to perfect absorption. Note that only the lumped elements
oriented parallel to the impinging $E$-field affect the resonance
of the unit-cell; for instance, when $(\theta,\varphi)=(75^{o},0)$
the TE (perpendicular) polarization is only affected by $RC_{y}$
elements and TM (parallel) polarization is only affected by $RC_{x}$
elements. Finally, our algorithm inherently assumes that the power
in the TE and TM polarizations is known (or can be measured) so that
the absorption coefficient can be translated to absorbed power. In
the context of the consensus workflow, we will assume that the possible
$RC_{y/x}$ values are discretized in a $10\times10$ grid covering
uniformly the axes span of Fig.~\ref{fig:Colormaps}.

The present physical implementation of the HSF can ABSense the direction
of any linear polarization in the principal planes ($xz$ and $yz$,
$\varphi=0$ or $90^{o}$, respectively), where the $x$- and $y$-oriented
lumped loads are decoupled and directly correspond to TE or TM polarizations;
ABSensing the direction of incidence of pure TE or TM (but not both)
polarizations, when $\varphi\in(0,90^{o})$ is also possible, but
requires for more careful cross-polarization coupling considerations
as both loads ($RC_{x/y}$) affect both polarizations (TE/TM) simultaneously.
The general case of ABSensing elliptical polarizations in a broad
frequency range and in the entire hemisphere, $\varphi\in(0,360^{o})$,
is a topic for future work requiring for more complicated `anisotropic'
unit-cell designs (optimized for higher resolution in $\varphi\theta$
vs. $RC$ measurement) and, additionally, the ability to measure the
current (or voltage) phase on the lumped elements.

Regarding the nano-node workflow parameters, we consider one nano-node
per HSF active element. Each nano-node is located exactly below the
corresponding element center and at $\nicefrac{t_{d}}{2}$ depth within
the substrate. In order to simulate the inter-nanonode communication
channel, we employ the model of \cite{NANOCOM2015Tsioliaridou,tsioliaridou2016stateless,Iyer.2009}
using the same physical-layer parameter values (frequency~$100$~GHz,
noise level~$0$~dBnW, SINR threshold~$-10$~dB, guard interval~$0.1$~nsec).
Assuming a bitrate of $100$~Gbps, we consider a consensus data packet
duration of approximately $1$~nsec (i.e., the $72$~bits of the
consensus packet plus preamble overheads rounding up at $100$~bits
total). The transmission power is set to 30~dBnW, yielding approximately~$20$
nodes within connectivity range.

Regarding the consensus process parameters, we set a TTM equal to
$50$~times the HSF operating period ($5$~GHz $\to0.2$~nsec),
to accommodate any transient EM phenomena (typically lasting $2-3$
periods) and obtain dependable average power flow values. The RD is
picked at random in the range $0$~to~$10$~packet duration(s)
to minimize collisions. The TTW is set to a marginally larger value
than the maximum RD, i.e., $12$~times the single packet duration.
Finally, we assign a $50$~\% consensus weight to the local optimum
of each node and an equal, $50$~\% aggregate weight to all incoming
consensus values.

\begin{figure}[t!]
\centering{}\includegraphics[width=1\linewidth]{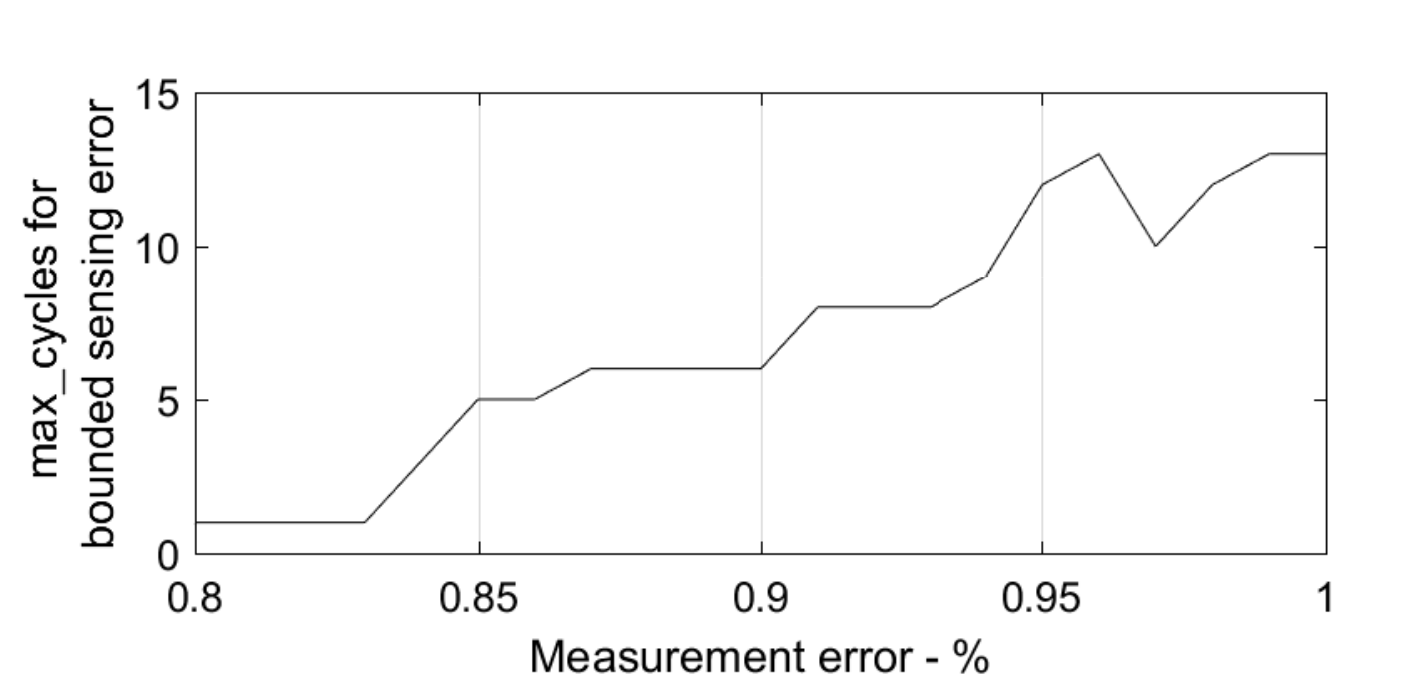} \caption{\label{fig:maxCycles}Consensus iterations required for mitigating
random errors in the nano-node power measurements.}
\end{figure}
In Fig.~\ref{fig:maxCycles} we consider the impinging EM wave of
Fig.~\ref{fig:Colormaps}(b). Additionally, we assume that the power
measurements of each node contain a fault expressed as a random percentage
over the actual value. We are interested in deducing the minimal max\_cycle
value that eliminates the error via the consensus value averaging.
The consensus process is shown to be very robust to such errors in
the measurements. For errors up to \textasciitilde$82$ ~\%, even
one consensus cycle is enough, while \textasciitilde$12${} cycles
can eliminate even the highest measurement errors.

\begin{figure}[t!]
\centering{}\includegraphics[width=1\linewidth]{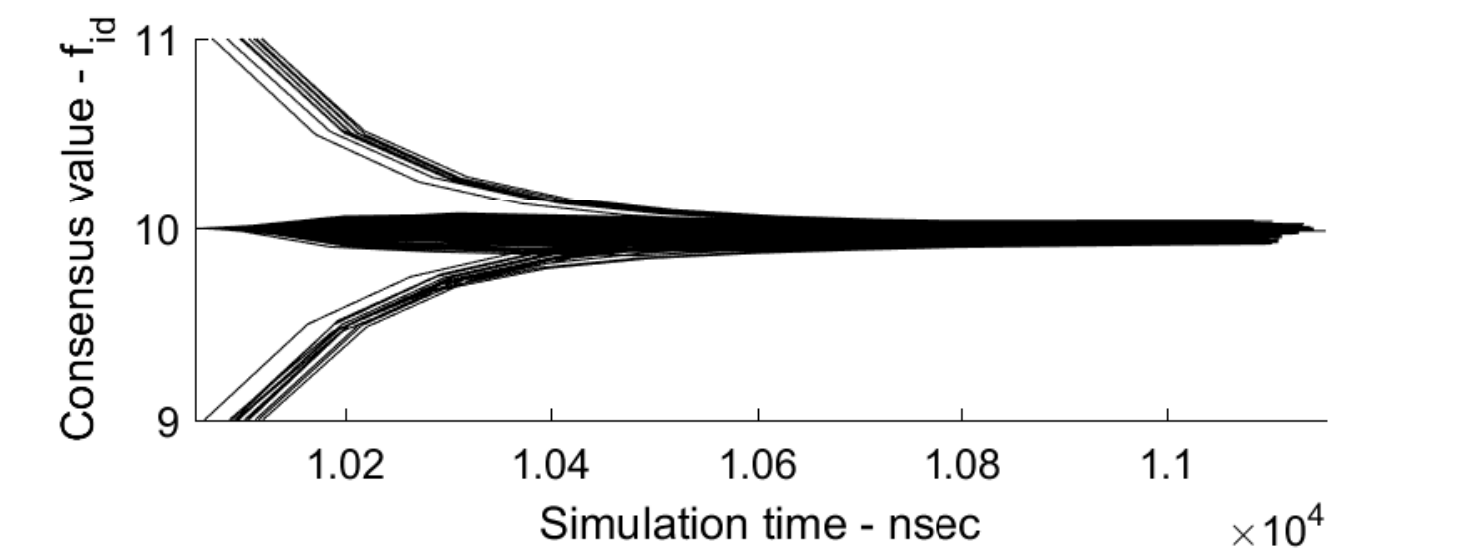} \caption{\label{fig:flows}Progression of each node's consensus (sensory) value
as the time progresses. }
\end{figure}
\begin{figure}[t!]
\begin{centering}
\includegraphics[width=1\linewidth]{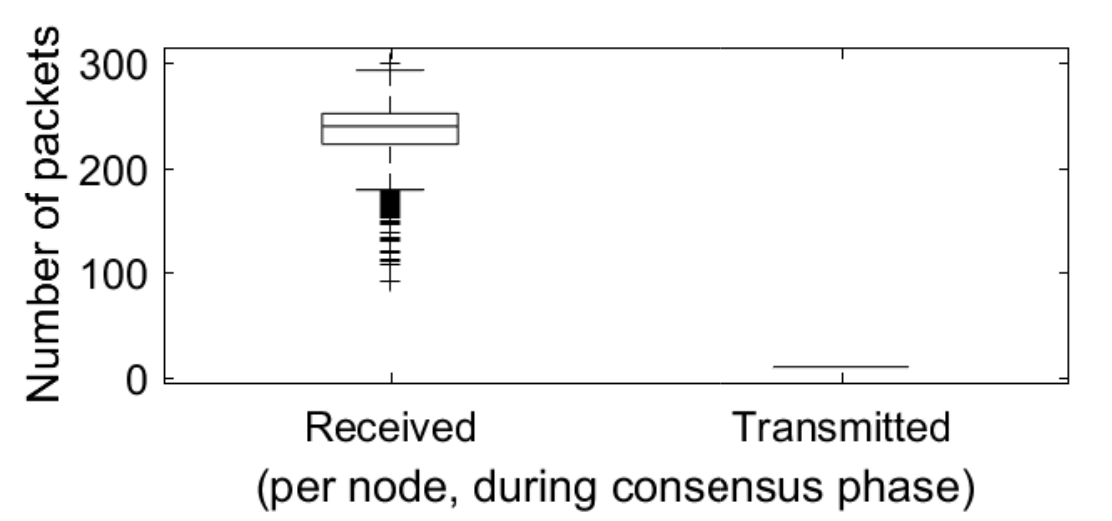}
\par\end{centering}
\centering{}\includegraphics[width=0.48\columnwidth]{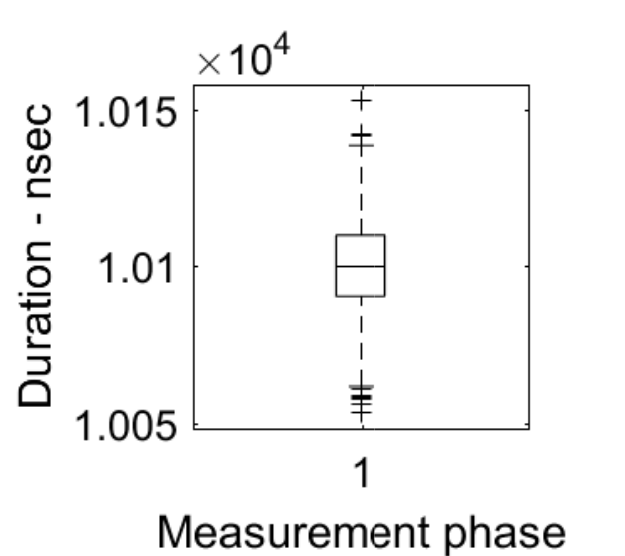}\includegraphics[width=0.48\columnwidth]{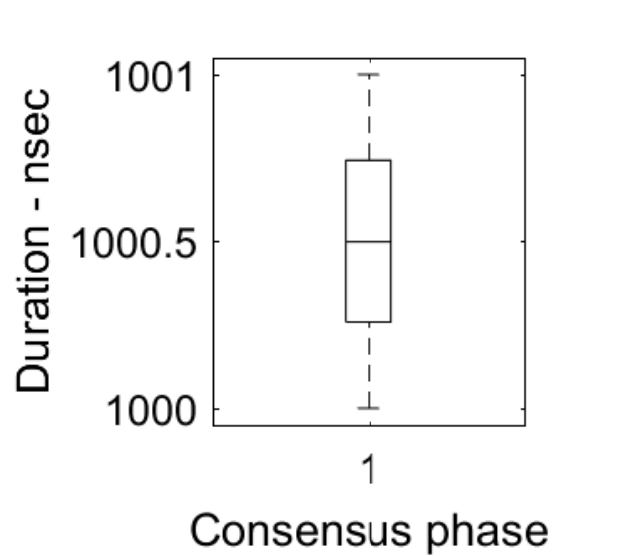}\caption{\label{fig:stats}Packet transmission statistics (per nano-node) during
consensus, and measurement/consensus phase durations.}
\end{figure}
In Fig.~\ref{fig:flows}-\ref{fig:stats} we focus on the $90$~\%
measurement error case and set max\_cycles to~$10$. Fig.~\ref{fig:flows}
plots the consensus (i.e., sensory) value progression versus time.
For ease of exposition, we employ the EM function identifier (cf.
rel.~(\ref{eq:example})) to denote the correct impinging wave attribute
($f_{id}=10$). As shown, the consensus process converges rapidly
even for nodes with completely erroneous initial measurement. Thus,
the consensus process is economic in terms of required packet transmissions.
The corresponding packet statistics are shown in Fig.~\ref{fig:stats}-top:
each node received $10$ packets (per cycle) from each of its $\sim20$
neighbors, subject to some losses due to collisions. The transmission
per node are strictly bounded and fully defined by the max\_cycles
value. Finally, as shown in Fig.~\ref{fig:stats}-bottom, the measurements
phase lasts for~$10\,\mu$sec (i.e., $10\times10$ RC values to iterate
over, times the time to obtain a single measurement), while the consensus
phase lasts for $1\,\mu$sec (i.e., max\_cyles times the RD and TTW).
At an aggregate EM wave sensing time of $11$~$\mu$sec, ABSense
shows promise for application in real-time-adapting HSFs.

\section{Conclusion and Future Work}

\label{sec:conclusion}

The manipulation and re-shaping of EM waves via intelligent metasurfaces
constitutes a key enabler of exotic capabilities in energy harvesting,
medical imaging and wireless communications. Nonetheless, accurate
EM manipulation requires the accurate sensing of the attributes (direction,
polarization, phase) of the wave impinging on a metasurface. Filling
in a gap in the related research, the study proposed a novel sensory
scheme that exploits the ambient intelligence and communication capabilities
of the HyperSurfaces, a novel meta-surface variant. An embedded nano-network
automatically tunes the HyperSurface to full absorb the impinging
waves, as exhibited by the increase of energy dissipating within it,
hence indirectly estimating its attributes. The proposed scheme, validated
via extensive EM simulations, can also constitute the basis for distributed
EM compiler processes, were the nano-network will auto-tune the HyperSurface
status to obtain any high-level objective in its macroscopic EM behavior.

In the future, the authors plan to extend the proposed scheme towards
sensing non-planar, complex EM waves, while also advancing the nano-network
consensus process into an always-on process that will run in parallel
with EM manipulation objectives.

\section{Acknowledgments}

This work was funded by the European Union via the Horizon 2020: Future
Emerging Topics call (FETOPEN), grant EU736876, project VISORSURF
(http://www.visorsurf.eu).

The authors would like to thank all members of the VISORSURF consortium
for the valuable, inter-disciplinary knowledge exchange that has taken
place since the project start.



\begin{thebibliography}{10}
\providecommand{\url}[1]{#1}
\csname url@samestyle\endcsname
\providecommand{\newblock}{\relax}
\providecommand{\bibinfo}[2]{#2}
\providecommand{\BIBentrySTDinterwordspacing}{\spaceskip=0pt\relax}
\providecommand{\BIBentryALTinterwordstretchfactor}{4}
\providecommand{\BIBentryALTinterwordspacing}{\spaceskip=\fontdimen2\font plus
\BIBentryALTinterwordstretchfactor\fontdimen3\font minus
  \fontdimen4\font\relax}
\providecommand{\BIBforeignlanguage}[2]{{%
\expandafter\ifx\csname l@#1\endcsname\relax
\typeout{** WARNING: IEEEtran.bst: No hyphenation pattern has been}%
\typeout{** loaded for the language `#1'. Using the pattern for}%
\typeout{** the default language instead.}%
\else
\language=\csname l@#1\endcsname
\fi
#2}}
\providecommand{\BIBdecl}{\relax}
\BIBdecl

\bibitem{yang2016programmable}
H.~Yang \emph{et~al.}, ``A programmable metasurface with dynamic polarization,
  scattering and focusing control,'' \emph{NPG Scientific reports}, vol.~6,
  2016.

\bibitem{PlasmonicMS}
H.~Wakatsuchi \emph{et~al.}, ``Waveform-dependent absorbing metasurfaces,''
  \emph{Physical Review Letters}, vol. 111, no.~24, 2013.

\bibitem{Tasolamprou2019720}
A.~Tasolamprou and othes, ``Experimental demonstration of ultrafast thz
  modulation in a graphene-based thin film absorber through negative
  photoinduced conductivity,'' \emph{ACS Photonics}, vol.~6, no.~3, 2019.

\bibitem{wallace2010analysis}
H.~Wallace, ``Analysis of rf imaging applications at frequencies over 100
  ghz,'' \emph{OSA Applied optics}, vol.~49, no.~19, 2010.

\bibitem{Liaskos:2018:UAS:3289258.3192336}
C.~Liaskos \emph{et~al.}, ``Using any surface to realize a new paradigm for
  wireless communications,'' \emph{Communications of the ACM}, vol.~61, no.~11,
  2018.

\bibitem{MSSurveyAllFunctionsAndTypes}
A.~Li \emph{et~al.}, ``Metasurfaces and their applications,''
  \emph{Nanophotonics}, vol.~7, no.~6, 2018.

\bibitem{AccessUPC}
S.~Abadal \emph{et~al.}, ``Computing and communications for the
  software-defined metamaterial paradigm: {A} context analysis,'' \emph{IEEE
  Access}, vol.~5, 2017.

\bibitem{Fieldsen0:online}
M.~electronics, ``Field sensors: montena technology sa,''
  \url{https://www.montena.com/system/pulse-measurement-fibre-optic-links/field-sensors/},
  (Accessed on 12/01/2018).

\bibitem{LiaskosComp}
C.~Liaskos \emph{et~al.}, ``Initial {UML} definition of the {HyperSurface}
  compiler middle-ware,'' \emph{European Commission, H2020-FETOPEN-2016-2017,
  Project VISORSURF: Accepted Public Deliverable D2.2, 31-Dec-2017, [Online:]
  \url{http://www.visorsurf.eu/m/VISORSURF-D2.2.pdf}}, 2017.

\bibitem{Diaz-Rubio:2017}
A.~D{\'\i}az-Rubio \emph{et~al.}, ``From the generalized reflection law to the
  realization of perfect anomalous reflectors,'' \emph{AAAS Sci. Adv.}, vol.~3,
  no.~8, 2017.

\bibitem{Tasolamprou201423147}
A.~Tasolamprou \emph{et~al.}, ``Experimentally excellent beaming in a two-layer
  dielectric structure,'' \emph{Opt. Express}, vol.~22, no.~19, 2014.

\bibitem{Niemi20133102}
T.~Niemi \emph{et~al.}, ``Synthesis of polarization transformers,'' \emph{IEEE
  Trans Antennas Propag}, vol.~61, no.~6, 2013.

\bibitem{Tsilipakos:2018}
O.~Tsilipakos \emph{et~al.}, ``Antimatched electromagnetic metasurfaces for
  broadband arbitrary phase manipulation in reflection,'' \emph{ACS Photonics},
  vol.~5, 2018.

\bibitem{Wang2018}
X.~Wang \emph{et~al.}, ``Extreme asymmetry in metasurfaces via evanescent
  fields engineering: Angular-asymmetric absorption,'' \emph{Phys Rev Lett},
  vol. 121, no.~25, 2018.

\bibitem{Cui:2014}
T.~J. Cui \emph{et~al.}, ``Coding metamaterials, digital metamaterials and
  programmable metamaterials,'' \emph{Light Sci. Appl.}, vol.~3, no.~10, 2014.

\bibitem{Huang:2017}
C.~Huang \emph{et~al.}, ``Reconfigurable metasurface for multifunctional
  control of electromagnetic waves,'' \emph{Adv. Opt. Mater.}, vol.~5, 2017.

\bibitem{Fu:2019}
F.~Liu \emph{et~al.}, ``Intelligent metasurfaces with continuously tunable
  local surface impedance for multiple reconfigurable functions,'' \emph{Phys.
  Rev. Appl.}, vol.~11, 2019.

\bibitem{DBLP:conf/WoWMoM/Liaskos}
C.~Liaskos \emph{et~al.}, ``Realizing wireless communication through
  software-defined hypersurface environments,'' in \emph{IEEE WoWMoM'18}.

\bibitem{DBLP:journals/corr/abs-1810-06329}
T.~Saeed \emph{et~al.}, ``Fault adaptive routing in metasurface controller
  networks,'' in \emph{NoCArc'18}.

\bibitem{Liaskos2019ADHOC}
C.~Liaskos \emph{et~al.}, ``A novel communication paradigm for high capacity
  and security via programmable indoor wireless environments in next generation
  wireless systems,'' \emph{Ad Hoc Networks}, vol.~87.

\bibitem{IEEEcomLiaskos}
------, ``A new wireless communication paradigm through software-controlled
  metasurfaces,'' \emph{IEEE Communications Magazine}, vol.~56, 2018.

\bibitem{Pelzman20161213}
C.~Pelzman and S.-Y. Cho, ``Plasmonic metasurface for simultaneous detection of
  polarization and spectrum,'' \emph{Opt. Lett.}, vol.~41, no.~6, 2016.

\bibitem{Chen2018110}
M.~Chen \emph{et~al.}, ``Detection of orbital angular momentum with metasurface
  at microwave band,'' \emph{IEEE Antennas Wirel. Propag. Lett.}, vol.~17,
  no.~1, 2018.

\bibitem{consensus1}
C.~Hadjicostis \emph{et~al.}, ``Robust distributed average consensus via
  exchange of running sums,'' \emph{IEEE Transactions on Automatic Control},
  vol.~61, no.~6, 2016.

\bibitem{consensus2}
L.~Tseng and N.~Vaidya, ``Fault-tolerant consensus in directed graphs,'' in
  \emph{ACM PODC'15}.

\bibitem{akram}
N.~Abuali \emph{et~al.}, ``Performance evaluation of routing protocols in
  electromagnetic nanonetworks,'' \emph{IEEE Access}, vol.~6, 2018.

\bibitem{NANOCOM2015Tsioliaridou}
A.~Tsioliaridou \emph{et~al.}, ``Corona: A coordinate and routing system for
  nanonetworks,'' in \emph{NANOCOM'15}.

\bibitem{julien}
H.~Mabed and J.~Bourgeois, ``A flexible medium access control protocol for
  dense terahertz nanonetworks,'' in \emph{NANOCOM'18}.

\bibitem{ICTsioliaridou}
A.~Tsioliaridou \emph{et~al.}, ``N3: Addressing and routing in 3d
  nanonetworks,'' in \emph{IEEE ICT'16}.

\bibitem{NANOCOMlocalizeSDM}
------, ``A novel protocol for network-controlled metasurfaces,'' in
  \emph{NANOCOM'17}.

\bibitem{al2018lagoon}
F.~Al-Turjman and K.~I. Kilic, ``Lagoon: a simple energy-aware routing protocol
  for wireless nano-sensor networks,'' \emph{IET Wireless Sensor Systems},
  2018.

\bibitem{ICCLiaskos}
C.~Liaskos \emph{et~al.}, ``A deployable routing system for nanonetworks,'' in
  \emph{IEEE ICC'16}.

\bibitem{estakhri2019inverse}
N.~M. Estakhri, B.~Edwards, and N.~Engheta, ``Inverse-designed metastructures
  that solve equations,'' \emph{AAAS Science}, vol. 363, no. 6433, 2019.

\bibitem{la2019curvilinear}
L.~La~Spada \emph{et~al.}, ``Curvilinear metasurfaces for surface wave
  manipulation,'' \emph{Scientific reports}, vol.~9, no.~1, 2019.

\bibitem{la2017near}
L.~La~Spada and L.~Vegni, ``Near-zero-index wires,'' \emph{OSA Optics express},
  vol.~25, no.~20, 2017.

\bibitem{vakil2011transformation}
A.~Vakil and N.~Engheta, ``Transformation optics using graphene,'' \emph{AAAS
  Science}, vol. 332, no. 6035, 2011.

\bibitem{Tasolamprou2018}
A.~Tasolamprou \emph{et~al.}, ``{Intercell wireless communication in
  software-defined metasurfaces},'' in \emph{ISCAS'18}, 2018.

\bibitem{Timoneda2018b}
X.~Timoneda \emph{et~al.}, ``{Channel Characterization for Chip-scale Wireless
  Communications within Computing Packages},'' in \emph{Proceedings of the NOCS
  '18}, 2018.

\bibitem{eharvest}
W.~Guo \emph{et~al.}, ``Multi-scale energy harvesting,'' in \emph{Wireless
  Information and Power Transfer: A New Paradigm for Green
  Communications}.\hskip 1em plus 0.5em minus 0.4em\relax Springer, 2018, pp.
  157--185.

\bibitem{CST}
\BIBentryALTinterwordspacing
{CST}, \emph{{Microwave Studio}}, 2016. [Online]. Available:
  \url{http://www.cst.com/}
\BIBentrySTDinterwordspacing

\bibitem{XJTechnologies.2013}
\BIBentryALTinterwordspacing
{XJ~Technologies}, \emph{{The AnyLogic Simulator}}, 2018. [Online]. Available:
  \url{http://www.anylogic.com/}
\BIBentrySTDinterwordspacing

\bibitem{tsioliaridou2016stateless}
A.~Tsioliaridou \emph{et~al.}, ``Stateless linear-path routing for 3d
  nanonetworks,'' in \emph{ACM NANOCOM'16}.

\bibitem{Iyer.2009}
A.~Iyer \emph{et~al.}, ``{What is the right model for wireless channel
  interference?}'' \emph{{IEEE Trans. Wir. Comm.}}, vol.~8, no.~5, 2009.

\end{thebibliography}
\end{document}